\documentclass[conference]{IEEEtran}

\usepackage[utf8]{inputenc}
\usepackage{amsmath,amsfonts,amssymb,bm}
\usepackage{cite}
\usepackage{xcolor}
\usepackage[nolist]{acronym}
\usepackage[innermargin=0.6in, outermargin=0.6in, top=0.6in, bottom=1in]{geometry}

\def\ve#1{{\mathchoice{\mbox{\boldmath$\displaystyle #1$}}%
		{\mbox{\boldmath$\textstyle #1$}}%
		{\mbox{\boldmath$\scriptstyle #1$}}%
		{\mbox{\boldmath$\scriptscriptstyle #1$}}}}

\newtheorem{theorem}{Theorem}
\newtheorem{definition}{Definition}

\newcommand{\Fqm}{\ensuremath{\mathbb F_{q^m}}}

\newcommand{\Fq}{\ensuremath{\mathbb F_{q}}}

\newcommand{\NN}{\ensuremath{\mathbb{N}}}
\newcommand{\set}[1]{\ensuremath{\mathcal{#1}}}

\newcommand{\aut}{\ensuremath{\theta}}
\newcommand{\autinv}{\ensuremath{\aut^{-1}}}

\newcommand{\SkewPolyringZeroDer}{\ensuremath{\Fqm[x;\aut]}}
\newcommand{\SkewPolyringZeroDerInv}{\ensuremath{\Fqm[x;\autinv]}}

\newcommand{\opev}[3]{\ensuremath{{#1}(#2)_{#3}}}
\newcommand{\opfull}[2]{\ensuremath{\mathcal{D}_{\aut,#1}(#2)}}

\newcommand{\opfullexp}[3]{\ensuremath{\mathcal{D}_{\aut,#1}^{#3}(#2)}}
\newcommand{\opfullexpinv}[3]{\ensuremath{\mathcal{D}_{\autinv,#1}^{#3}(#2)}}

\DeclareMathOperator{\mpol}{mpol}
\newcommand{\minpolyOpNoX}[2]{\ensuremath{\mpol_{#1}^{#2}}}
\newcommand{\minpolyOp}[2]{\ensuremath{\minpolyOpNoX{#1}{#2}(x)}}

\newcommand{\OCompl}[1]{\ensuremath{\mathcal{O}({#1})}}

\newcommand{\defeq}{:=}
\newcommand{\eqdef}{=:}

\renewcommand{\bar}{\overline}

\newcommand{\modl}{\; \mathrm{mod}_\mathrm{l} \;}
\newcommand{\modr}{\; \mathrm{mod}_\mathrm{r} \;}

\DeclareMathOperator{\wt}{wt}
\DeclareMathOperator{\rk}{rk}

\newcommand{\mystack}[2]{\ensuremath{\genfrac{}{}{0pt}{}{#1}{#2}}}

\renewcommand{\vec}[1]{\ve{#1}} 
\newcommand{\mat}[1]{\ensuremath{\bm{#1}}}

\newcommand{\opVandermonde}[3]{\ensuremath{\mat{V}_{\aut}^{#1}(#2)_{#3}}}
\newcommand{\opMoore}[3]{\ensuremath{\mathfrak{M}_{\aut}^{#1}(#2)_{#3}}}
\newcommand{\opMooreInv}[3]{\ensuremath{\mathfrak{M}_{\autinv}^{#1}(#2)_{#3}}}
\newcommand{\genNorm}[2]{\ensuremath{\mathcal{N}_{\aut}^{#1}(#2)}}
\newcommand{\genNormInv}[2]{\ensuremath{\mathcal{N}_{\autinv}^{#1}(#2)}}

\newcommand{\lclm}{\ensuremath{\mathrm{lclm}}}

\renewcommand{\a}{\vec{a}}

\renewcommand{\c}{\vec{c}}
\newcommand{\e}{\vec{e}}

\newcommand{\s}{\vec{s}}

\newcommand{\x}{\vec{x}}
\newcommand{\y}{\vec{y}}

\newcommand{\A}{\mat{A}}
\newcommand{\B}{\mat{B}}
\newcommand{\C}{\mat{C}}

\newcommand{\E}{\mat{E}}
\newcommand{\G}{\mat{G}}
\newcommand{\I}{\mat{I}}
\renewcommand{\H}{\mat{H}}

\newcommand{\X}{\mat{X}}
\newcommand{\Y}{\mat{Y}}
\newcommand{\0}{\ensuremath{\mathbf 0}}
\newcommand{\vecalpha}{\ensuremath{\boldsymbol{\alpha}}}
\newcommand{\vecbeta}{\ensuremath{\boldsymbol{\beta}}}

\newcommand{\vecxi}{\ensuremath{\boldsymbol{\xi}}}
\newcommand{\nVec}{\ensuremath{\vec{n}}}

\newcommand{\mycode}[1]{\ensuremath{\mathcal{#1}}}

\newcommand{\linRS}[1]{\ensuremath{\mathrm{LRS}[#1]}}

\newcommand{\SumRankWeight}{\ensuremath{\wt_{\Sigma R}}}

\newcommand{\SumRankDist}{d_{\ensuremath{\Sigma}R}}

\newcommand{\SumRankWeightWPartition}[1]{\SumRankWeight^{#1}}
\newcommand{\SumRankDistWPartition}[1]{\SumRankDist^{#1}}

\newcommand{\numbFullErrors}{t_{\indFullErrors}}
\newcommand{\numbRowErasures}{t_{\indRowErasures}}
\newcommand{\numbColErasures}{t_{\indColErasures}}
\newcommand{\numbErrors}{\tau}

\newcommand{\numbFullErrorsInBlock}[1]{\numbFullErrors^{(#1)}}
\newcommand{\numbRowErasuresInBlock}[1]{\numbRowErasures^{(#1)}}
\newcommand{\numbColErasuresInBlock}[1]{\numbColErasures^{(#1)}}
\newcommand{\numbErrorsInBlock}[1]{\numbErrors^{(#1)}}

\newcommand{\indFullErrors}{F}
\newcommand{\indRowErasures}{R}
\newcommand{\indColErasures}{C}

\newcommand{\indErrorType}{T}
\newcommand{\numbErrorType}{t_{\indErrorType}}
\newcommand{\numbErrorTypeInBlock}[1]{\numbErrorType^{(#1)}}

\newcommand{\eFullErrors}{\e_\indFullErrors}
\newcommand{\eRowErasures}{\e_\indRowErasures}
\newcommand{\eColErasures}{\e_\indColErasures}
\newcommand{\eErrorType}{\e_\indErrorType}

\newcommand{\autInvXiWIndex}[1]{\widetilde{\xi_{#1}}} 
\newcommand{\autInvXiInvWIndex}[1]{\widehat{\xi_{#1}}} 

\newcommand{\myspace}[1]{\mathcal{#1}}

\newcommand{\insertions}{\ensuremath{\gamma}}

\newcommand{\deletions}{\ensuremath{\delta}}

\newcommand{\txSpace}{\ensuremath{\myspace{V}}}
\newcommand{\rxSpace}{\ensuremath{\myspace{U}}}

\newcommand{\txSpaceVec}{\ensuremath{\vec{\txSpace}}}
\newcommand{\rxSpaceVec}{\ensuremath{\vec{\rxSpace}}}

\newcommand{\shot}[2]{\ensuremath{{#1}^{(#2)}}}

\newcommand{\insertionsShot}[1]{\ensuremath{\shot{\insertions}{#1}}}
\newcommand{\deletionsShot}[1]{\ensuremath{\shot{\deletions}{#1}}}

\newcommand{\txSpaceShot}[1]{\ensuremath{\shot{\txSpace}{#1}}}
\newcommand{\rxSpaceShot}[1]{\ensuremath{\shot{\rxSpace}{#1}}}

\newcommand{\pe}{\ensuremath{\gamma}}

\newcommand{\shots}{\ensuremath{\ell}}

\newcommand{\skewrev}[1]{\ensuremath{\bar{#1}}}

\newcommand{\ELP}{\ensuremath{\lambda}}
\newcommand{\ELPfull}{\ensuremath{\ELP_{\indFullErrors}}}
\newcommand{\ELProw}{\ensuremath{\ELP_{\indRowErasures}}}
\newcommand{\ELPcol}{\ensuremath{\ELP_{\indColErasures}}}
\newcommand{\ELPcolWIndex}[1]{\ensuremath{\ELP_{\indColErasures, #1}}}

\newcommand{\ELPcolRev}{\ensuremath{\skewrev{\ELPcol}}}
\newcommand{\ELPcolRevWIndex}[1]{\ensuremath{\skewrev{\ELPcolWIndex{#1}}}}

\newcommand{\ELPfullcol}{\ensuremath{\ELP_{\indFullErrors \indColErasures}}}

\newcommand{\ESP}{\ensuremath{\sigma}}
\newcommand{\ESPfull}{\ensuremath{\ESP_\indFullErrors}}
\newcommand{\ESProw}{\ensuremath{\ESP_\indRowErasures}}
\newcommand{\ESPcol}{\ensuremath{\ESP_\indColErasures}}

\newcommand{\ESProwRev}{\ensuremath{\skewrev{\ESProw}}}

\newcommand{\ESPfullrow}{\ensuremath{\ESP_{\indFullErrors \indRowErasures}}}
\newcommand{\ESPfullrowWIndex}[1]{\ensuremath{\ESP_{\indFullErrors \indRowErasures, #1}}}

\newcommand{\ESPsyndrome}{s_{RC}}
\newcommand{\ELPsyndrome}{s_{CR}}
\newcommand{\coeffpower}[2]{\mathfrak{c}_{#2}(#1)}

\begin{acronym}
	\acro{BCH}{Bose--Chaudhuri--Hocquenghem}
	\acro{BMD}{bounded minimum distance}
	\acro{ESP}{error span polynomial}
	\acro{ELP}{error locator polynomial}
	\acro{GMD}{generalized minimum distance}
	\acro{SRS}{skew Reed--Solomon}
	\acro{ISRS}{interleaved skew Reed--Solomon}
	\acro{lclm}{least common left multiple}
	\acro{LRS}{linearized Reed--Solomon}
	\acro{LLRS}{lifted linearized Reed--Solomon}
	\acro{ILRS}{interleaved linearized Reed--Solomon}
	\acro{LILRS}{lifted interleaved linearized Reed--Solomon}
	\acro{MDS}{maximum distance separable}
	\acro{MSRD}{maximum sum-rank distance}
	\acro{MSD}{maximum skew distance}
	\acro{RS}{Reed--Solomon}
	\acro{LEEA}{linearized extended Euclidean algorithm}
	\acro{SEEA}{skew extended Euclidean algorithm}
\end{acronym}

\IEEEoverridecommandlockouts

\begin{document}

\allowdisplaybreaks
\setlength{\abovedisplayskip}{5pt}
\setlength{\belowdisplayskip}{5pt}

\title{
    Error-Erasure Decoding of Linearized Reed--Solomon Codes in the Sum-Rank Metric
}

\author{
    \IEEEauthorblockN{Felicitas Hörmann, Hannes Bartz}
    \IEEEauthorblockA{
        \textit{Institute of Communications and Navigation} \\
        \textit{German Aerospace Center (DLR)}\\
        Oberpfaffenhofen, Germany \\
        \{felicitas.hoermann, hannes.bartz\}@dlr.de
    }
    \and
    \IEEEauthorblockN{Sven Puchinger
        \thanks{
            This work was done while S. Puchinger was with the Department of Applied Mathematics and Computer Science, Technical University of Denmark (DTU), Lyngby, Denmark and the Department of Electrical and Computer Engineering, Technical University of Munich, Munich, Germany.
            Within this period he was supported by the European Union’s Horizon 2020 research and innovation program under the Marie Sklodowska-Curie grant agreement no. 713683 and by the European Research Council (ERC) under the European Union’s Horizon 2020 research and innovation programme (grant agreement no. 801434).
        }
    }
    \IEEEauthorblockA{
        \textit{Hensoldt Sensors GmbH} \\
        Ulm, Germany \\
        sven.puchinger@tum.de
    }
}

\maketitle

\begin{abstract}
    Codes in the sum-rank metric have various applications in error control for multishot network coding, distributed storage and code-based cryptography.
    Linearized Reed--Solomon (LRS) codes contain Reed--Solomon and Gabidulin codes as subclasses and fulfill the Singleton-like bound in the sum-rank metric with equality.
    We propose the first known error-erasure decoder for LRS codes to unleash their full potential for multishot network coding by incorporating erasures into the known syndrome-based Berlekamp--Massey-like decoder.
    This allows to correct $\numbFullErrors$ full errors, $\numbRowErasures$ row erasures and $\numbColErasures$ column erasures up to $2\numbFullErrors + \numbRowErasures + \numbColErasures \leq n-k$ in the sum-rank metric requiring at most $\OCompl{n^2}$ operations in $\Fqm$, where $n$ is the code's length and $k$ its dimension.
    We show how the proposed decoder can be used to correct errors in the sum-subspace metric that occur in (noncoherent) multishot network coding.

\end{abstract}


\section{Introduction}\label{sec:introduction}

The sum-rank metric is a generalization of both the Hamming and the rank metric and was first considered in~\cite[Sec.~III]{lu2005unified} for designing space-time codes.
Later, N{\'o}brega and Uch{\^o}a-Filho showed that the sum-rank metric is suitable for error control in \emph{coherent} multishot network coding and proposed a multilevel code construction~\cite{nobrega2010multishot}.
Other constructions of codes in the sum-rank metric include partial unit memory codes constructed from rank-metric codes~\cite{wachter2011partial,wachter2015convolutional}, convolutional codes~\cite{napp2017mrd,napp2018faster} and variable block-size constructions~\cite{byrne2021fundamental}.

Mart{\'\i}nez-Pe{\~n}as introduced \ac{LRS} codes which include Reed--Solomon and Gabidulin codes as special cases~\cite{martinez2018skew}.
\ac{LRS} codes fulfill the Singleton-like bound in the sum-rank metric with equality and thus are \ac{MSRD} codes.
The interest in \ac{LRS} and other sum-rank metric codes keeps increasing as they have multiple widespread applications as e.g. multishot network coding~\cite{nobrega2010multishot,martinez2019reliable}, locally repairable codes~\cite{martinez2019Universal}, space-time codes~\cite{lu2005unified} and code-based quantum-resistant cryptography~\cite{puchinger2020generic}.
Recently, it was shown that \emph{interleaved}~\cite{bartz2021decoding,bartz2022fast} and \emph{folded}~\cite{hormann2021efficient} variants of LRS codes can be decoded beyond the unique decoding radius.
The concept of row and column erasures, i.e. the partial knowledge of the column and row space of the error, respectively, was generalized from the rank metric~\cite{GabidulinParamonovTretjakov_RankErasures_1991} to the sum-rank metric in~\cite{puchinger2020generic}.

In this paper, we extend the syndrome-based error-only decoder for \ac{LRS} codes from~\cite{puchinger2021private} to a Berlekamp--Massey-like error-erasure decoder that can correct $\numbFullErrors$ full errors, $\numbRowErasures$ row and $\numbColErasures$ column erasures as long as $2\numbFullErrors + \numbRowErasures + \numbColErasures \leq n-k$ requiring $\OCompl{n^2}$ operations in $\Fqm$, where $n$ denotes the length and $k$ the dimension of the code, respectively.
The proposed algorithm is inspired by the error-erasure decoding algorithms for Gabidulin codes from~\cite{richter2004ErrorAndErasure,silva2009error}.
Further, we show how the results can be used to decode lifted \ac{LRS} codes for error control in multishot network coding~\cite{silva_rank_metric_approach}.

Up to our knowledge, the proposed decoder is the first decoding scheme for \ac{LRS} codes that is capable of correcting both errors and row/column erasures in the sum-rank metric.

\section{Notation and Preliminaries}\label{sec:notation-and-preliminaries}

For a prime power $q$ and a positive integer $m$, let $\Fq$ denote a finite field of order $q$ and $\Fqm \supseteq \Fq$ its extension field with extension degree $m$.
Under a fixed basis of $\Fqm$ over $\Fq$ there is a bijection between any element $a\in\Fqm$ and a length-$m$ column vector $\a$ over $\Fq$.
Recall further that an element $\pe \in \Fqm$ is called \emph{primitive} in $\Fqm$ if it generates $\Fqm^{\ast} \defeq \Fqm \setminus \{0\}$.

Consider an automorphism $\aut: \Fqm \to \Fqm$.
Two elements $a,b \in \Fqm$ are called \emph{$\aut$-conjugate,} if there exists an element $c \in \Fqm^{\ast}$ such that $b = \aut(c)ac^{-1}$.
The \emph{conjugacy class} $\set{C}(a)$ is the set of all $\aut$-conjugates of $a$ and $\set{C}(0)$ is called trivial conjugacy class.
Note that $\aut$-conjugacy defines an equivalence relation on $\Fqm$ and that the conjugacy classes form a partition of $\Fqm$ (see e.g.~\cite{lam1988vandermonde}).
Let $\{\xi_1, \dots, \xi_{\shots}\} \subseteq \Fqm^{\ast}$ be a set of representatives of distinct nontrivial conjugacy classes of $\Fqm$.
Then, $\set{C}(\aut(\xi_1)), \dots, \set{C}(\aut(\xi_{\shots}))$ as well as $\set{C}(\aut^{-1}(\xi_1^{-1})), \dots, \set{C}(\aut^{-1}(\xi_{\shots}^{-1}))$ are distinct and nontrivial classes because they are injective images of $\set{C}(\xi_1), \dots, \set{C}(\xi_{\shots})$.

\subsection{Skew Polynomials}

For an automorphism $\aut$ on $\Fqm$, the non-commutative \emph{skew polynomial ring} $\SkewPolyringZeroDer$ (with zero derivation) consists of all formal polynomials $\sum_i f_i x^{i-1}$ having finitely many nonzero coefficients $f_i \in \Fqm$.
It is equipped with ordinary polynomial addition and the multiplication is determined by $x f_i = \aut(f_i) x$ for all $f_i \in \Fqm$.
Naturally, the \emph{degree} of a nonzero skew polynomial $f(x) = \sum_i f_i x^{i-1}$ is $\deg(f) \defeq \max \{i: f_{i+1} \neq 0\}$ whereas the degree of the zero polynomial is set to $-\infty$.
We use the notation $\SkewPolyringZeroDer_{<k} \defeq \{f \in \SkewPolyringZeroDer: \deg(f) < k\}$ for any $k \geq 0$.

Note that $\SkewPolyringZeroDer$ is a left and right Euclidean ring which ensures the existence of $q_l, r_l, q_r, r_r \in \SkewPolyringZeroDer$ such that
\begin{align}
    a &= q_l b + r_l \quad &\text{with } &\deg(r_l) < \deg(b)
    \\
    \text{and} \qquad a &= b q_r + r_r \quad &\text{with } &\deg(r_r) < \deg(b)
\end{align}
for every $a,b \in \SkewPolyringZeroDer$.
We write $r_l = a \modl b$ and $r_r = a \modr b$, respectively.

The product $p = f \cdot g \in \SkewPolyringZeroDer$ of two skew polynomials $f, g \in \SkewPolyringZeroDer$ with $d_f \defeq \deg(f)$ and $d_g \defeq \deg(g)$ has degree $d_f + d_g$.
The coefficients $p_l$ of $p$ with $\min(d_f, d_g) + 1 \leq l \leq \max(d_f, d_g) + 1$ can be computed as (see~\cite{li2014transform})
\begin{align}
    p_l &= \sum_{i=1}^{d_f+1}f_i\aut^{i-1}(g_{l-i+1}) \qquad \text{if } d_f \leq d_g
    \label{eq:skew_product_coeffs}
    \\
    \text{and} \qquad p_l &= \sum_{i=1}^{d_g+1}f_{l-i+1}\aut^{l-i}(g_{i}) \qquad \text{if } d_g \leq d_f.
    \label{eq:skew_product_coeffs_rev}
\end{align}

The \emph{(partial) $\aut$-reverse} of $f$ with respect to an integer $t \geq d_f$ is defined as $\skewrev{f}(x) = \sum_{i=1}^{t+1} \skewrev{f}_i x^{i-1}$,
where $\skewrev{f}_i = \aut^{i-t-1}(f_{t-i+2})$ for all $i=1,\dots,t+1$~\cite[p.~574]{li2014transform},~\cite[Sec.~2.4]{silva2009error}.

For all $a \in \Fqm$, the \emph{generalized power function} is defined as $\genNorm{0}{a} = 1$ and as $\genNorm{i}{a}=\aut^{i-1}(a) \cdot \genNorm{i-1}{a}$ for all $i > 0$ (see~\cite{lam1988vandermonde}).
This notion is used to define the operator
\begin{equation}\label{eq:def_op}
    \opfull{a}{b} \defeq \aut(b)a
    \quad \text{for all } a,b \in \Fqm
\end{equation}
and for all $i \in \NN$ and $a,b \in \Fqm$ its powers (see~\cite[Prop.~32]{martinez2018skew})
\begin{equation}\label{eq:def_op_exp}
    \opfullexp{a}{b}{i}
    =\opfull{a}{\opfullexp{a}{b}{i-1}}
    =\aut^i(b)\genNorm{i}{a}.
\end{equation}

A vector $\nVec \defeq (n_1,\dots,n_\shots) \in \NN^{\shots}$ is called a \emph{length partition} of $n \in \NN$ if $n = \sum_{i=1}^{\shots} n_i$.
We divide $\x \in \Fqm^n$ into $\shots$ blocks with respect to $\nVec$ by writing $\x = (\x^{(1)} \mid \dots \mid \x^{(\shots)})$ with $\x^{(i)} \in \Fqm^{n_i}$.
For a fixed ordered $\Fq$-basis of $\Fqm$ there are isomorphisms $\Fqm^{n_i} \to \Fq^{m \times n_i}$ which allow to define the $\Fq$-rank of each vector $\x^{(i)}$, i.e. $\rk_{\Fq}(\x^{(i)})$, as the rank of the corresponding matrix.
The \emph{generalized Moore matrix} for $\x$, a vector $\a = (a_1, \ldots, a_\shots) \in \Fqm^{\shots}$ and a parameter $d \in \NN^{\ast}$ is defined as
\begin{equation}\label{eq:def_gen_moore_mat}
    \opMoore{d}{\x}{\a} \defeq
    \begin{pmatrix}
        \opVandermonde{d}{\x^{(1)}}{a_1} & \cdots & \opVandermonde{d}{\x^{(\shots)}}{a_\shots}
    \end{pmatrix} \in \Fqm^{d \times n}
\end{equation}
where its Vandermonde-like submatrices $\opVandermonde{d}{\x^{(i)}}{a_i}$ are
\begin{equation}
    \opVandermonde{d}{\x^{(i)}}{a_i} \defeq
    \begin{pmatrix}
        x^{(i)}_1 & \cdots & x^{(i)}_{n_i}
        \\
        \opfull{a_i}{x^{(i)}_1} & \cdots & \opfull{a_i}{x^{(i)}_{n_i}}
        \\[-4pt]
        \vdots & \ddots & \vdots
        \\
        \opfullexp{a_i}{x^{(i)}_1}{d-1} & \cdots & \opfullexp{a_i}{x^{(i)}_{n_i}}{d-1}
    \end{pmatrix}
\end{equation}
for $1 \leq i \leq \shots$.
If $\a$ contains representatives of pairwise distinct nontrivial conjugacy classes of $\Fqm$ and $\rk_{\Fq}(\x^{(i)}) = n_i$ for all $1 \leq i \leq \shots$, we have by~\cite[Thm.~2]{martinez2018skew} and~\cite[Thm.~4.5]{lam1988vandermonde} that $\rk_{\Fqm}(\opMoore{d}{\x}{\a}) = \min(d, n)$.

The \emph{generalized operator evaluation} of a skew polynomial $f \in \SkewPolyringZeroDer$ at an element $b \in \Fqm$ with respect to an evaluation parameter $a \in \Fqm$ is defined as (see~\cite{leroy1995pseudolinear,martinez2018skew})
\begin{equation}\label{eq:def_gen_op_eval}
  \opev{f}{b}{a}=\sum_{i}f_i\opfullexp{a}{b}{i-1}.
\end{equation}
For a fixed evaluation parameter $a$ the generalized operator evaluation forms an $\Fq$-linear map~\cite{leroy1995pseudolinear}.
The evaluation of a product of two skew polynomials $f,g \in \SkewPolyringZeroDer$ satisfies
$\opev{(f\cdot g)}{b}{a} = \opev{f}{\opev{g}{b}{a}}{a}$ for all $a,b \in \Fqm$~\cite{martinez2019private}.

The \emph{minimal (skew) polynomial} that vanishes on the set $\{b_1^{(i)}, \dots, b_{n_i}^{(i)}\} \subseteq \Fqm$ with respect to the evaluation parameter $a_i \in \Fqm$ for all $i = 1, \dots, \shots$ is defined as
\begin{equation}
    \opev{\minpolyOpNoX{\{a_i\}_{i=1}^{\shots}}{\{b_{\kappa}^{(i)}\}_{\kappa=1}^{n_i}}}{b_{\kappa}^{(i)}}{a_i} = 0
    \quad \text{for all} \quad
    \begin{aligned}
        1 &\leq \kappa \leq n_i
        \\
        1 &\leq i \leq \shots
    \end{aligned}
    .
\end{equation}
When assuming $b_{\kappa}^{(i)} \neq 0$ for all indices it can be computed as
\begin{equation}
    \label{eq:min_poly}
    \minpolyOp{\{a_i\}_{i=1}^{\shots}}{\{b_{\kappa}^{(i)}\}_{\kappa=1}^{n_i}} = \lclm\left(x-\frac{\aut(b_{\kappa}^{(i)})a_i}{b_{\kappa}^{(i)}}\right)_{\mystack{1 \leq \kappa \leq n_i}{1 \leq i \leq \shots}}
\end{equation}
where $\lclm(\cdot)$ denotes the \acl{lclm} of the polynomials in the bracket~\cite[Sec.~1.3.1]{caruso2019residues}.
Its degree is at most $\sum_{i=1}^{\shots} n_i$ and equality holds if and only if the $b_{\kappa}^{(i)}$ belonging to the same evaluation parameter $a_i$ are $\Fq$-linearly independent and the evaluation parameters $a_i$ are representatives of different nontrivial conjugacy classes of $\Fqm$.

\subsection{Sum-Rank Metric and Linearized Reed--Solomon Codes}

The \emph{sum-rank weight} of a vector $\x \in \Fqm^n$ with respect to the length partition $\nVec \in \NN^{\shots}$ is
\begin{equation}
    \label{eq:sum-rank_weight}
    \SumRankWeightWPartition{\nVec}(\x) = \sum_{i=1}^{\shots} \rk_{\Fq}(\x^{(i)}).
\end{equation}
The metric introduced by
$\SumRankDistWPartition{\nVec}(\x, \y) \defeq \SumRankWeightWPartition{\nVec}(\x - \y)$ for all $\x, \y \in \Fqm^n$
is called the \emph{sum-rank metric} (with respect to $\nVec$).
When $\nVec$ is clear from the context, we simply write $\SumRankWeight$ and $\SumRankDist$, respectively.

An $[n,k]$ \emph{linear sum-rank metric code} $\mycode{C} \subseteq \Fqm^n$ is defined as a $k$-dimensional $\Fqm$-linear subspace of $\Fqm^n$ and thus has length $n$.
Its \emph{minimum sum-rank distance} is
\begin{align}
    \SumRankDist(\mycode{C})
    &\defeq \min_{\substack{\x, \y \in \mycode{C},\\ \x \neq \y}} \{ \SumRankDist(\x, \y)\}
    = \min_{\substack{\x \in \mycode{C}, \\ \x\neq \0}} \{ \SumRankWeight(\x)\},
\end{align}
where the last equality follows by linearity.
Codes achieving the Singleton-like bound $\SumRankDist(\mycode{C}) \leq n-k+1$ (see e.g.~\cite[Prop. 34]{martinez2018skew}) with equality are called \emph{\acf{MSRD} codes}.

\begin{definition}[Linearized Reed--Solomon Codes]\label{def:LRS_codes}
    Let $\vecxi=(\xi_1,\dots,\xi_\shots)\in\Fqm^\shots$ be a vector containing representatives of pairwise distinct nontrivial conjugacy classes of $\Fqm$ and consider a length partition $\nVec := (n_1,\dots\allowbreak,n_\shots)\in\NN^\shots$ of $n \in \NN$.
    Let the vectors $\vecbeta^{(i)}=(\beta_1^{(i)},\dots,\beta_{n_i}^{(i)}) \in \Fqm^{n_i}$ contain $\Fq$-linearly independent elements of $\Fqm$ for all $i=1,\dots,\shots$ and define $\vecbeta \defeq \left(\vecbeta^{(1)}\mid\dots\mid\vecbeta^{(\shots)}\right) \in \Fqm^n$.
    A \emph{\acf{LRS} code} $\linRS{\aut,\vecbeta,\vecxi,\shots;\nVec,k} \subseteq \Fqm^{n}$ of length $n$ and dimension $k$ is defined as
    \begin{equation}
        \left\{
        \left(
        \opev{f}{\vecbeta^{(1)}}{\xi_1} \mid \dots \mid \opev{f}{\vecbeta^{(\shots)}}{\xi_\shots}
        \right): f\in\SkewPolyringZeroDer_{<k}\right\}
    \end{equation}
    where $\opev{f}{\vecbeta^{(i)}}{\xi_i} \defeq (\opev{f}{\beta_1^{(i)}}{\xi_i}, \dots, \opev{f}{\beta_{n_i}^{(i)}}{\xi_i})$ for $1 \leq i \leq \shots$.
\end{definition}

\ac{LRS} codes have minimum sum-rank distance $n-k+1$ and are thus~\ac{MSRD}~\cite[Thm. 4]{martinez2018skew}.
Furthermore, $\linRS{\aut,\vecbeta,\vecxi,\shots;\nVec,k}$ has a generator matrix of the form $\G = \opMoore{k}{\vecbeta}{\vecxi}$~\cite[Sec.~3.3]{martinez2018skew}.

The \emph{dual} of an \ac{LRS} code can be described as (see~\cite{caruso2019residues,caruso2021duals})
\begin{equation}
    \linRS{\aut,\vecbeta,\vecxi,\shots;\nVec,k}^\perp=\linRS{\autinv,\vecalpha,\autinv(\vecxi),\shots;\nVec,n-k}
\end{equation}
where the vector $\vecalpha=(\vecalpha^{(1)}\mid\dots\mid\vecalpha^{(\shots)})\in\Fqm^n$ (with $\vecalpha^{(i)} = (\alpha_1^{(i)}, \dots, \alpha_{n_i}^{(i)}) \allowbreak \in \Fqm^{n_i}$ for $i = 1, \dots, \shots$) satisfies
\begin{equation}\label{eq:syndrome_eqs_lrs}
    \sum_{i=1}^{\shots}\sum_{\kappa=1}^{n_i}\alpha_{\kappa}^{(i)}\opfullexp{\xi_i}{\beta_{\kappa}^{(i)}}{l-1}=0
    \quad \text{for all } l=1,\dots,n-1
\end{equation}
and has sum-rank weight $\SumRankWeight(\vecalpha)=n$ (see~\cite[Thm.~4]{martinez2019reliable}).
Hence, there exists a parity-check matrix of $\linRS{\aut,\vecbeta,\vecxi,\shots;\nVec,k}$ of the form $\H = \opMooreInv{n-k}{\vecalpha}{\autinv(\vecxi)}$.

\section{Error-Erasure Decoding}\label{sec:error-erasure-decoding}

\subsection{Channel Model}\label{subsec:error-erasure-channel}

We consider an additive sum-rank channel with fixed error weight $\numbErrors \in \NN$ and incorporate three types of errors.
Next to $\numbFullErrors$ conventional (full) errors, we allow $\numbRowErasures$ \emph{row erasures} whose column spaces are known and $\numbColErasures$ \emph{column erasures} whose row spaces are given by the channel.
Erasures and the notions of row and column support in the sum-rank metric have already been studied in~\cite{puchinger2020generic} and naturally generalize the respective rank-metric concepts (see e.g.~\cite{GabidulinParamonovTretjakov_RankErasures_1991, gabidulin2008errorAndErasure, silva_rank_metric_approach}).

The error vector $\e = (\e^{(1)} \mid \dots \mid \e^{(\shots)}) \in \Fqm^n$ is assumed to have sum-rank weight $\SumRankWeight(\e) = \numbErrors = \numbFullErrors + \numbRowErasures + \numbColErasures$.
To emphasize in which block the errors occurred, we write $\numbErrorsInBlock{i} = \rk_{\Fq}(\e^{(i)}) = \numbFullErrorsInBlock{i} + \numbRowErasuresInBlock{i} + \numbColErasuresInBlock{i}$ for all $i=1,\dots,\shots$.
In this context, the transmission of a codeword $\c\in\linRS{\aut,\vecbeta,\vecxi,\shots;\nVec,k}$ yields a channel observation $\y \in \Fqm^n$ of the form $\y = \c + \e$.

We have already implicitly assumed that $\e$ has an additive decomposition $\e = \eFullErrors + \eRowErasures + \eColErasures$ with respect to the considered error types, where $\e_\indErrorType \in \Fqm^n$ satisfies $\SumRankWeight(\e_\indErrorType) = \numbErrorType$ for all $\indErrorType \in \{\indFullErrors, \indRowErasures, \indColErasures\}$.
Application of~\cite[Lem.~5]{puchinger2020generic} leads for all error types $\indErrorType \in \{\indFullErrors, \indRowErasures, \indColErasures\}$ to a representation
\begin{align}
    \eErrorType &=
    \underbrace{(\a_\indErrorType^{(1)}\mid\dots\mid\a_\indErrorType^{(\shots)})}_{\eqdef \a_\indErrorType \in \Fqm^{\numbErrorType}}
    \cdot
    \underbrace{
    \setlength{\arraycolsep}{1pt}
    \renewcommand{\arraystretch}{0.6}
    \begin{pmatrix}
     \B_\indErrorType^{(1)} \\[-5pt]
     & \ddots \\
     & & \B_\indErrorType^{(\shots)}
    \end{pmatrix}
    }_{\eqdef \B_\indErrorType \in \Fq^{\numbErrorType \times n}}
\end{align}
where both $\a_\indErrorType^{(i)} \in \Fqm^{\numbErrorTypeInBlock{i}}$ and $\B_\indErrorType^{(i)} \in \Fq^{\numbErrorTypeInBlock{i} \times n_i}$ have rank $\numbErrorTypeInBlock{i}$ for all $i = 1, \dots, \shots$.
Note that the entries of $\a_\indErrorType^{(i)}$ form a basis of the column space of $\e_\indErrorType^{(i)}$ and the rows of $\B_\indErrorType^{(i)} = (b_{\indErrorType, j, \kappa}^{(i)})_{j, \kappa}$ are a basis of its row space.
Hence, according to the definition of row and column erasures, $\a_\indRowErasures$ and $\B_\indColErasures$ are known to the receiver.

We define the \emph{error locators} corresponding to the $i$-th block of $\e$ for each $i = 1, \dots, \shots$ as the $\numbErrorsInBlock{i}$ components of the vector
$\x^{(i)} = (\x_\indFullErrors^{(i)}, \x_\indRowErasures^{(i)}, \x_\indColErasures^{(i)}) \in \Fqm^{\numbErrorsInBlock{i}}$,
where $\x_\indErrorType^{(i)} \in \Fqm^{\numbErrorTypeInBlock{i}}$ with $\indErrorType \in \{\indFullErrors, \indRowErasures, \indColErasures\}$ has the entries
\begin{equation}
    x_{\indErrorType, j}^{(i)} \defeq \sum_{\kappa=1}^{n_i} b_{\indErrorType, j, \kappa}^{(i)} \alpha_{\kappa}^{(i)}
    \quad \text{for all }
    j = 1, \dots, \numbErrorTypeInBlock{i}.
\end{equation}
For simplicity we renumber the entries of $\x^{(i)}$ and reference them as $x_r^{(i)}$ for $1 \leq r \leq \numbErrorsInBlock{i}$ and $1 \leq i \leq \shots$ in the following.
Similarly we write $\a^{(i)} = (\a_\indFullErrors^{(i)}, \a_\indRowErasures^{(i)}, \a_\indColErasures^{(i)}) \in \Fqm^{\numbErrorsInBlock{i}}$ for the vector containing the \emph{error values} of the $i$-th error block and $a_r^{(i)}$ for its entries (for $r = 1, \dots, \numbErrorsInBlock{i}$ and $i = 1, \dots, \shots$).

Now consider the syndrome $\s = \y\H^\top = \e\H^\top$.
Then, the entries of $\s$ can be written as
\begin{align}
    \label{eq:syndrome_entries}
    s_l
    &= \sum_{i=1}^{\shots} \sum_{r=1}^{\numbErrorsInBlock{i}} a_r^{(i)} \opfullexpinv{\autinv(\xi_i)}{x_r^{(i)}}{l-1}
\end{align}
for all $l = 1, \dots, n-k$.
When letting $\x = (\x^{(1)} \mid \dots \mid \x^{(\shots)})$ and $\a = (\a^{(1)} \mid \dots \mid \a^{(\shots)})$ denote the vectors containing all error locators and all error values, respectively, we have the equivalent formulation (see also~\cite{puchinger2020generic})
\begin{equation}\label{eq:lrs_syndrome_system_err_val}
    \X\a^\top = \s^\top
    \quad \text{ with } \X = \opMooreInv{n-k}{\x}{\autinv(\vecxi)} \in \Fqm^{(n-k) \times \numbErrors}
\end{equation}
where $\autinv(\vecxi)$ is defined as $(\autinv(\xi_1), \dots, \autinv(\xi_\shots))$.

By applying $\aut^{l-1}$ to~\eqref{eq:syndrome_entries} we get
\begin{align}
    \aut^{l-1}(s_l)
    &=\sum_{i=1}^{\shots}\sum_{r=1}^{\numbErrorsInBlock{i}}x_r^{(i)}\opfullexp{\xi_i}{a_r^{(i)}}{l-1}\label{eq:lrs_rev_syndrome_short}
\end{align}
for all $l = 1, \dots, n-k$ and equivalently
\begin{equation}\label{eq:lrs_syndrome_system_err_loc}
    \A\x^\top = \widetilde{\s}^\top
    \quad \text{ with } \A = \opMoore{n-k}{\a}{\vecxi} \in \Fqm^{(n-k) \times \numbErrors}
\end{equation}
and $\widetilde{\s}=(s_1,\aut(s_2),\dots,\aut^{n-k-1}(s_{n-k}))\in\Fqm^{n-k}$.

\subsection{ESP and ELP Key Equation}

We can now define the \ac{ESP} $\ESP \in \SkewPolyringZeroDerInv$ as the minimal polynomial
\begin{equation}
    \ESP(x) = \sum_{\nu=1}^{\numbErrors+1} \ESP_\nu x^{\nu-1}
    \quad \text{with} \quad
    \opev{\ESP}{a_{r}^{(i)}}{\autinv(\xi_i^{-1})} = 0
\end{equation}
for all $r=1,\dots,\numbErrorsInBlock{i}$ and all $i = 1, \dots, \shots$.
In an analogous manner the \ac{ELP} $\ELP \in \SkewPolyringZeroDerInv$ is the minimal polynomial
\begin{equation}
    \ELP(x) = \sum_{\nu=1}^{\numbErrors+1} \ELP_\nu x^{\nu-1}
    \quad \text{with} \quad
    \opev{\ELP}{x_{r}^{(i)}}{\autinv(\xi_i)} = 0
\end{equation}
for every $r=1,\dots,\numbErrorsInBlock{i}$ and $i = 1, \dots, \shots$.
Note that both $\ESP$ and $\ELP$ are members of a skew polynomial ring with respect to the \emph{inverse} automorphism $\autinv$ and we consider generalized operator evaluation parameters that are different compared to the \ac{LRS} code construction.

Let us now express the \ac{ESP} and the \ac{ELP} as products of three polynomials related to the different error types.
This will prove beneficial for incorporating the knowledge about row and column erasures into the decoder.
We write
\begin{align}
    \ESP(x) &= \ESPcol(x) \cdot \ESPfull(x) \cdot \ESProw(x)
    \\
    \text{and } \quad \ELP(x) &= \ELProw(x) \cdot \ELPfull(x) \cdot \ELPcol(x)
\end{align}
where the partial \ac{ESP}s $\ESPfull, \ESProw, \ESPcol \in \SkewPolyringZeroDerInv$ are defined as
$\ESP_\indErrorType(x) = \sum_{\nu=1}^{\numbErrorType+1} \ESP_{\indErrorType, \nu} x^{\nu-1}$
for $\indErrorType \in \{\indFullErrors, \indRowErasures, \indColErasures\}$ being the minimal polynomials satisfying
\begin{align}
    \opev{(\ESPcol \cdot \ESPfull \cdot \ESProw)}{a_{\indColErasures,j}^{(i)}}{\autinv(\xi_i^{-1})} &= 0  \quad \text{for } j=1,\dots,\numbColErasuresInBlock{i},
    \notag \\
    \opev{(\ESPfull \cdot \ESProw)}{a_{\indFullErrors,j}^{(i)}}{\autinv(\xi_i^{-1})} &= 0  \quad \text{for } j=1,\dots,\numbFullErrorsInBlock{i},
    \label{eq:def_part_esp_w_erasures} \\
    \text{and} \quad \opev{\ESProw}{a_{\indRowErasures,j}^{(i)}}{\autinv(\xi_i^{-1})} &= 0  \quad \text{for } j=1,\dots,\numbRowErasuresInBlock{i}
    \notag
\end{align}
for all $i=1,\dots,\shots$, respectively.
Similarly, the partial error locator polynomials $\ELPfull, \ELProw, \ELPcol \in \SkewPolyringZeroDerInv$ with
$\ELP_\indErrorType(x) = \sum_{\nu=1}^{\numbErrorType+1} \ELP_{\indErrorType, \nu} x^{\nu-1}$
for $\indErrorType \in \{\indFullErrors, \indRowErasures, \indColErasures\}$ are given as the minimal polynomials that satisfy
\begin{align}
    \opev{(\ELProw \cdot \ELPfull \cdot \ELPcol)}{x_{\indRowErasures,j}^{(i)}}{\autinv(\xi_i)} &= 0  \quad \text{for } j=1,\dots,\numbRowErasuresInBlock{i},
    \notag \\
    \opev{(\ELPfull \cdot \ELPcol)}{x_{\indFullErrors,j}^{(i)}}{\autinv(\xi_i)} &= 0 \quad \text{for } j=1,\dots,\numbFullErrorsInBlock{i},
    \label{eq:def_part_elp_w_erasures} \\
    \text{and} \quad \opev{\ELPcol}{x_{\indColErasures,j}^{(i)}}{\autinv(\xi_i)} &= 0 \quad \text{for } j=1,\dots,\numbColErasuresInBlock{i}
    \notag
\end{align}
for all $1 \leq i \leq \shots$, respectively.
Note that since $\a_\indRowErasures$ and $\B_\indColErasures$ are known, we can compute $\ESProw$ and $\ELPcol$ using~\eqref{eq:min_poly}.

Let $s \in \SkewPolyringZeroDerInv$ be the \emph{syndrome polynomial} that is obtained from the syndrome $\s = (s_1, \dots, s_{n-k})$ as
\begin{equation}
   s(x) = \sum_{l=1}^{n-k} s_l x^{l-1}.
\end{equation}
Consider the auxiliary syndrome polynomial
\begin{equation}
    \ESPsyndrome(x) \defeq \ESProw(x) \cdot s(x) \cdot \ELPcolRev(x) \in \SkewPolyringZeroDerInv
\end{equation}
where $\ELPcolRev$ denotes the $\autinv$-reverse of $\ELPcol$ with respect to $\numbColErasures$.
This allows to derive the \ac{ESP} key equation that is the main ingredient of (the \ac{ESP} variant of) our error-erasure decoder.

\begin{theorem}[ESP Key Equation]\label{thm:esp_key_equation_w_erasures}
    There is a skew polynomial $\omega \in \SkewPolyringZeroDerInv$ of degree less than $\numbErrors = \numbFullErrors + \numbRowErasures + \numbColErasures$ such that
    \begin{equation}\label{eq:esp_key_equation}
        \ESPfull(x) \cdot \ESPsyndrome(x) \equiv \omega(x) \modr x^{n-k}.
    \end{equation}
\end{theorem}

\begin{IEEEproof}
    Let us write $\ESPfullrow \defeq \ESPfull \cdot \ESProw$ as well as $\autInvXiWIndex{i} \defeq \autinv(\xi_i)$ and $\autInvXiInvWIndex{i} \defeq \autinv(\xi_i^{-1})$ ($i = 1, \dots, \shots$) for brevity.
    For $\numbFullErrors + \numbRowErasures + 1 \leq l \leq n-k$ the $l$-th coefficient of $\ESPfull \cdot \ESProw \cdot s$ is
    \begin{align}
        (\ESPfull \cdot \ESProw \cdot s)_l
        &= \sum_{\nu=1}^{\numbFullErrors+\numbRowErasures+1} \ESPfullrowWIndex{\nu} \aut^{-(\nu-1)}(s_{l-\nu+1})
        \notag \\
        &= \sum_{i=1}^{\shots} \sum_{r=1}^{\numbErrorsInBlock{i}} \opev{\ESPfullrow}{a_r^{(i)}}{\autInvXiInvWIndex{i}} \opfullexpinv{\autInvXiWIndex{i}}{x_r^{(i)}}{l-1}.
        \label{eq:ESP_proof_other_LRS_decoding_problem}
    \end{align}
    For $\numbColErasures + 1 \leq l \leq \numbFullErrors + \numbRowErasures + n - k$ we have
    \begin{align*}
        &(\ESPfull \cdot \ESProw \cdot s \cdot \ELPcolRev)_l
        = \sum_{\nu=1}^{\numbColErasures+1} (\ESPfullrow \cdot s)_{l-\nu+1} \aut^{-(l-\nu)}(\ELPcolRevWIndex{\nu})
        \\
        &= \sum_{i=1}^{\shots} \sum_{r=1}^{\numbErrorsInBlock{i}} \opev{\ESPfullrow}{a_r^{(i)}}{\autInvXiInvWIndex{i}} \aut^{-(l-\numbColErasures-1)}(\genNorm{l-\numbColErasures-1}{\xi_i}) \cdot
        \\
        &\phantom{=}~ \cdot \aut^{-(l-\numbColErasures-1)} \left( \sum_{\nu=1}^{\numbColErasures+1} \ELPcolWIndex{\numbColErasures-\nu+2} \opfullexpinv{\autInvXiWIndex{i}}{x_r^{(i)}}{\numbColErasures-\nu+1}\right)
        \\
        &= \sum_{i=1}^{\shots} \sum_{r=1}^{\numbErrorsInBlock{i}} \opev{\ESPfullrow}{a_r^{(i)}}{\autInvXiInvWIndex{i}} \aut^{-(l-\numbColErasures-1)}(\genNorm{l-\numbColErasures-1}{\xi_i}) \cdot
        \\
        &\phantom{=}~ \cdot \aut^{-(l-\numbColErasures-1)}\left( \opev{\ELPcol}{x_r^{(i)}}{\autInvXiWIndex{i}} \right)
        = 0
    \end{align*}
    where the second equality follows from
    \begin{equation}
        \genNormInv{l-\nu}{\autInvXiWIndex{i}}
        = \aut^{-(l-\numbColErasures-1)} \Big( \genNormInv{\numbColErasures-\nu+1}{\autInvXiWIndex{i}} \cdot \genNorm{l-\numbColErasures-1}{\xi_i} \Big).
    \end{equation}
    Since $\ESPfull \cdot \ESProw \cdot s \cdot \ELPcolRev = \ESPfull \cdot \ESPsyndrome$, the proof is complete.
\end{IEEEproof}

For the ELP variant of the decoder, we exploit a different auxiliary syndrome $\ELPsyndrome \in \SkewPolyringZeroDerInv$.
It is defined as
\begin{equation}
    \ELPsyndrome(x) = \ELPcol(x) \cdot \skewrev{s}(x) \cdot \coeffpower{\ESProwRev}{n-k-1}
\end{equation}
where $\coeffpower{\ESProwRev}{n-k-1} \defeq \aut^{n-k-1}(\ESProwRev(\aut^{-(n-k-1)}(x)))$ denotes the polynomial obtained from $\ESProwRev$ by applying $\aut^{n-k-1}$ to all its coefficients.
Moreover, $\skewrev{s}$ is the $\autinv$-reverse of the syndrome polynomial $s$ with respect to $n-k-1$.
We obtain the following key equation.

\begin{theorem}[ELP Key Equation]\label{thm:elp_key_equation_w_erasures}
    There is a $\psi \in \SkewPolyringZeroDerInv$ having degree less than $\numbErrors = \numbFullErrors + \numbRowErasures + \numbColErasures$ that satisfies
    \begin{equation}\label{eq:elp_key_equation}
        \ELPfull(x) \cdot \ELPsyndrome(x) \equiv \psi(x) \modr x^{n-k}.
    \end{equation}
\end{theorem}

\begin{IEEEproof}[Sketch of Proof]
    Let us write $\ELPfullcol \defeq \ELPfull \cdot \ELPcol$ and compute
    \begin{align}
        (\ELPfullcol \cdot \skewrev{s})_l
        &= \sum_{i=1}^{\shots} \sum_{r=1}^{\numbErrorsInBlock{i}} \opev{\ELPfullcol}{x_r^{(i)}}{\autinv(\xi_i)} \opfullexp{\xi_i}{a_r^{(i)}}{n-k-l}
        \label{eq:ELP_proof_other_LRS_decoding_problem}
    \end{align}
    for $\numbFullErrors + \numbColErasures + 1 \leq l \leq n-k$.
    Similar to the ESP variant, we can exploit that $\opev{\ESProw}{a_r^{(i)}}{\autinv(\xi_i^{-1})} = 0$ and finally obtain $(\ELPfull \cdot \ELPsyndrome)_l = 0$ for all $l = \numbRowErasures + 1, \dots, \numbFullErrors + \numbColErasures + d - 1$.
\end{IEEEproof}

Observe that both key equations can be expressed as a homogeneous system of $n-k-\numbErrors$ linear equations in $\numbFullErrors$ variables.
Similar arguments as in~\cite[p.~132]{gabidulin1992AFastMatrixDecodingAlg} combined with~\cite[Thm.~1.3.7]{caruso2019residues} imply that its coefficient matrix has full rank.
Hence, a unique solution exists if and only if $n-k-\numbErrors \geq \numbFullErrors$, that is if $2\numbFullErrors + \numbRowErasures + \numbColErasures \leq n-k$.
As we will see shortly, this is the only necessary constraint on the number of errors and erasures and therefore the decoding radius of our decoder.

\subsection{The Decoding Algorithm}

Suppose we receive a vector $\y=\c+\e \in \Fqm^n$ with $\SumRankWeight(\e)=\numbErrors = \numbFullErrors + \numbRowErasures + \numbColErasures$ along with the side-information $\a_{R}$ for the $\numbRowErasures$ row erasures and $\B_{C}$ for the $\numbColErasures$ column erasures from the channel.
Then our decoder proceeds as follows:
\begin{enumerate}
    \item Compute the syndrome $\s = \y \H^{\top}$ and the syndrome polynomial $s(x) = \sum_{l=1}^{n-k} s_l x^{l-1} \in \SkewPolyringZeroDerInv$.
    \item Compute the error locators $x_{\indColErasures,j}^{(i)} = \sum_{\kappa=1}^{n_i}b_{\indColErasures,j,\kappa}^{(i)}\alpha_{\kappa}^{(i)}$ for all $j=1,\dots,\numbColErasuresInBlock{i}$.
    \item Compute the skew polynomials
          \begin{equation*}
              \ELP_C = \minpolyOpNoX{\{\autinv(\xi_i^{-1})\}_{i=1}^{\shots}}{\{x_{C,j}^{(i)}\}_{j=1}^{\numbColErasuresInBlock{i}}}
              \text{ and }
              \ESP_R = \minpolyOpNoX{\{\autinv(\xi_i^{-1})\}_{i=1}^{\shots}}{\{a_{R,j}^{(i)}\}_{j=1}^{\numbRowErasuresInBlock{i}}}.
          \end{equation*}
    \item Recover $\x$ and $\a$ by using one of the two variants:
    \\
    \textbf{ESP Variant}
        \begin{enumerate}
            \item Compute the auxiliary syndrome $\ESPsyndrome(x) = \ESProw(x) \cdot s(x) \cdot \ELPcolRev(x)$.
            \item Recover $\ESPfull$ by solving the ESP key equation~\eqref{eq:esp_key_equation}.
            \item Find $\Fq$-linearly independent $a_{\indFullErrors,1}^{(i)},\dots,a_{\indFullErrors,\numbFullErrorsInBlock{i}}^{(i)}$ such that $\opev{\ESPfull}{a_{\indFullErrors,j}^{(i)}}{\autinv(\xi_i^{-1})}=0$ for all $j=1,\dots,\numbFullErrorsInBlock{i}$.
            \item Solve the \ac{LRS} syndrome decoding problem~\eqref{eq:ESP_proof_other_LRS_decoding_problem} to get $\opev{(\ESPfull \cdot \ESProw)}{a_{\indColErasures,j}^{(i)}}{\autinv(\xi_i^{-1})}$ for $j = 1, \dots, \numbColErasuresInBlock{i}$.
                  Namely, solve $\hat{\s} = \hat{\e} \hat{\H}^{\top}$ with $\hat{\H} = \opMooreInv{n-k-\numbFullErrors-\numbRowErasures}{\hat{\vecalpha}}{\autinv(\vecxi)}$, where $\hat{\alpha}_r^{(i)} = \opfullexpinv{\autinv(\xi_i)}{x_r^{(i)}}{\numbFullErrors+\numbRowErasures}$ for $i = 1, \dots, \shots$ and $r = 1, \dots, \numbErrorsInBlock{i}$, and with $\hat{\e}$ having sum-rank weight at most $\numbColErasures$.
            \item Compute $\ESPcol = \minpolyOpNoX{\{\autinv(\xi_i^{-1})\}_{i=1}^{\shots}}{\{\opev{(\ESPfull \cdot \ESProw)}{a_{\indColErasures,j}^{(i)}}{\autinv(\xi_i^{-1})}\}_{j=1}^{\numbColErasuresInBlock{i}}}$ and $\ESP = \ESPcol \cdot \ESPfull \cdot \ESProw$.
            \item Find $\Fq$-linearly independent $a_1^{(i)},\dots,a_{\numbErrorsInBlock{i}}^{(i)}$ such that $\opev{\ESP}{a_r^{(i)}}{\autinv(\xi_i^{-1})}=0$ for all $r=1,\dots,\numbErrorsInBlock{i}$, and $i=1,\dots,\shots$.
            \item Solve~\eqref{eq:lrs_syndrome_system_err_loc} for $\x$.
        \end{enumerate}
        \textbf{ELP Variant}
        \begin{enumerate}
            \item Compute $\ELPsyndrome(x) = \ELPcol(x) \cdot \skewrev{s}(x) \cdot \coeffpower{\ESProwRev}{n-k-1}$.
            \item Recover $\ELPfull$ by solving the ELP key equation~\eqref{eq:elp_key_equation}.
            \item Find $\Fq$-linearly independent $x_{\indFullErrors,1}^{(i)},\dots,x_{\indFullErrors,\numbFullErrorsInBlock{i}}^{(i)}$ such that $\opev{\ELPfull}{x_{\indFullErrors,j}^{(i)}}{\autinv(\xi_i)}=0$ for all $j=1,\dots,\numbFullErrorsInBlock{i}$ and $i=1,\dots,\shots$.
            \item Solve the \ac{LRS} syndrome decoding problem~\eqref{eq:ELP_proof_other_LRS_decoding_problem} to obtain $\opev{(\ELPfull \cdot \ELPcol)}{x_{\indRowErasures,j}^{(i)}}{\autinv(\xi_i^{-1})}$ for all $i = 1, \dots, \shots$ and $j = 1, \dots, \numbRowErasuresInBlock{i}$.
            \item Compute $\ELProw = \minpolyOpNoX{\{\autinv(\xi_i^{-1})\}_{i=1}^{\shots}}{\{\opev{(\ELPfull \cdot \ELPcol)}{x_{\indRowErasures,j}^{(i)}}{\autinv(\xi_i)}\}_{j=1}^{\numbRowErasuresInBlock{i}}}$ and $\ELP = \ELProw \cdot \ELPfull \cdot \ELPcol$.
            \item Find $\Fq$-linearly independent $x_1^{(i)},\dots,x_{\numbErrorsInBlock{i}}^{(i)}$ such that $\opev{\ELP}{x_r^{(i)}}{\autinv(\xi_i)}=0$ for all $r=1,\dots,\numbErrorsInBlock{i}$ and $i=1,\dots,\shots$.
            \item Solve~\eqref{eq:lrs_syndrome_system_err_val} for $\a$.
        \end{enumerate}
    \item Recover $\B_T^{(i)}$ from $\x_T^{(i)}$ for $\indErrorType \in \{\indFullErrors, \indRowErasures\}$ and $i = 1, \dots, \shots$.
          Namely, compute the $j$-th row of $\B_T^{(i)}$ using a left inverse of the $\Fq$-expansion of $\vecalpha^{(i)}$ and the expansion of $x_{T,j}^{(i)}$ for $\indErrorType \in \{\indFullErrors, \indRowErasures\}$, $i = 1, \dots, \shots$, and $j = 1, \dots, \numbErrorTypeInBlock{i}$.
    \item Compute $\e = \a_{\indFullErrors} \cdot \B_{\indFullErrors} + \a_{\indRowErasures} \cdot \B_{\indRowErasures} + \a_{\indColErasures} \cdot \B_{\indColErasures}$ and return $\c = \y - \e$.
\end{enumerate}

The complexity-dominating tasks in the proposed error-erasure decoding algorithm can be accomplished as follows.
All involved minimal polynomials have at most $n$ roots and can hence be computed recursively in $\OCompl{n^2}$ operations in $\Fqm$ using~\eqref{eq:min_poly}.
The key equations~\eqref{eq:esp_key_equation} and~\eqref{eq:elp_key_equation} as well as the systems~\eqref{eq:lrs_syndrome_system_err_val} and~\eqref{eq:lrs_syndrome_system_err_loc} can be solved via skew feedback shift register synthesis with complexity $\OCompl{(n-k)^2}$~\cite{Sidorenko2011SkewFeedback}.
The appearing syndrome decoding problems can be solved e.g. with a generalized version of Gabidulin's rank-metric decoder from~\cite[Sec.~6]{Gabidulin_TheoryOfCodes_1985}, which we will present in detail in an extended version of this paper, and complexity $\OCompl{\numbErrors^2}$.
The root spaces of a skew polynomial $f \in \SkewPolyringZeroDer$ with respect to different evaluation parameters $a_1, \dots, a_\shots$ are uniquely determined according to~\cite[Prop.~1.3.7]{caruso2019residues}.
Bases for these root spaces can be computed by using the method from~\cite[Chap.~11.1]{berlekamp2015algebraic} for each evaluation parameter, requiring at most $\OCompl{n^3}$ operations in $\Fq$ or $\OCompl{n^2}$ operations in $\Fqm$, respectively.
Overall, the proposed error-erasure decoder has complexity $\OCompl{n^2}$ operations in $\Fqm$.

\begin{theorem}[Error-Erasure Decoding]
    Consider an \ac{LRS} code $\linRS{\aut,\vecbeta,\vecxi,\shots;\nVec,k}$.
    If the number of full errors $\numbFullErrors$, of row erasures $\numbRowErasures$ and of column erasures $\numbColErasures$ satisfies $2 \numbFullErrors + \numbRowErasures + \numbColErasures \leq n-k$, then the proposed decoder can recover the transmitted codeword requiring at most $\OCompl{n^2}$ operations in $\Fqm$.
\end{theorem}

We verified the results for the proposed error-erasure decoder by a proof-of-concept implementation in SageMath~\cite{stein_sagemath}.

\section{Applications}

In~\cite{silva2009error} and~\cite{silva_rank_metric_approach} it was shown that the decoding problem for constant-dimension codes in the subspace metric can be cast to an error and row/column erasure decoding problem in the rank metric.
By combining the ideas from~\cite{martinez2019reliable},~\cite{silva2009error} and~\cite{silva_rank_metric_approach}, our error-erasure decoder can be used to decode \emph{lifted} \ac{LRS} codes for error control in (noncoherent) multishot random linear network coding with respect to the sum-subspace metric~\cite{nobrega2009multishot}.
For an \ac{LRS} code $\mycode{C}=\linRS{\aut,\vecbeta,\vecxi,\shots;\nVec,k}$ the \emph{lifted} \ac{LRS} code $\mathcal{I}(\mycode{C})$ consists of all lifted codewords obtained by blockwise application of the lifting operation from~\cite[Def.~3]{silva_rank_metric_approach} (see~\cite{martinez2019reliable}).
Namely, the lifting of $\c = (\c^{(1)}\mid\dots\mid\c^{(\shots)})\in\mycode{C}$ is given by
\begin{equation}
    \mathcal{I}(\c) \defeq \left( \langle (\I_{n_1}, \C^{(1)\top}) \rangle_{\Fq}, \dots, \langle (\I_{n_\shots}, \C^{(\shots)\top})\rangle_{\Fq} \right)
\end{equation}
where $\C^{(i)}\in\Fq^{m\times n_i}$ is the expansion of $\c^{(i)}\in\Fqm^{n_i}$ over $\Fq$ for all $i=1,\dots,\shots$ and $\langle \cdot \rangle_{\Fq}$ denotes the $\Fq$-linear row space of a matrix.
After the transmission of a tuple $\txSpaceVec = (\txSpaceShot{1}, \dots, \txSpaceShot{\shots}) \in\mathcal{I}(\mycode{C})$ over a multishot operator channel~\cite{nobrega2009multishot} with overall $\insertions=\sum_{i=1}^{\shots}\insertionsShot{i}$ insertions and $\deletions=\sum_{i=1}^{\shots}\deletionsShot{i}$ deletions we receive a tuple $\rxSpaceVec=(\rxSpaceShot{1},\dots,\rxSpaceShot{\shots})$.
By applying the reduction~\cite[Def.~4]{silva_rank_metric_approach} to each received component space $\rxSpaceShot{i}$ we get $\Y^{(i)}=\C^{(i)}+\E^{(i)}\in\Fq^{m\times n_i}$ where $\E^{(i)}$ can be decomposed with respect to the different error types as described in Section~\ref{subsec:error-erasure-channel} and has $\Fq$-rank $\numbErrorsInBlock{i}=\numbFullErrorsInBlock{i} + \numbRowErasuresInBlock{i} + \numbColErasuresInBlock{i}$.
Therefore, the decoding problem for lifted \ac{LRS} codes in the sum-subspace metric reduces to an error and row/column erasure decoding problem in the sum-rank metric, which can be solved by the proposed decoder.
Similar as for one-shot subspace codes we have that $\insertionsShot{i}=\numbFullErrorsInBlock{i}+\numbRowErasuresInBlock{i}$ and $\deletionsShot{i}=\numbFullErrorsInBlock{i}+\numbColErasuresInBlock{i}$ (see e.g.~\cite{bartz2017algebraic}) where $\numbColErasuresInBlock{i}$ is also referred to as the number of \emph{erasures} and $\numbRowErasuresInBlock{i}$ is also referred to as the number of \emph{deviations} (see~\cite{silva_rank_metric_approach,silva2009error}).

Thus, the proposed decoder can correct an overall number of insertions and deletions up to $\insertions+\deletions\leq n-k$, which coincides with the decoding region of the decoders from~\cite{martinez2019reliable},~\cite{bartz2021decoding} and~\cite{bartz2022fast}.

Other applications of the presented error-erasure decoder include e.g. \ac{GMD}-inspired randomized decoding algorithms for cryptography (see e.g.~\cite{WeakKeysFL,RandomizedGabDecoding}) as well as error-erasure decoding problems in the sum-rank metric arising in any context.

\section{Conclusion}\label{sec:conclusion}

We presented a Berlekamp--Massey-like error-erasure decoder for \acf{LRS} codes that can correct $\numbFullErrors$ full errors, $\numbRowErasures$ row erasures and $\numbColErasures$ column erasures up to $2\numbFullErrors + \numbRowErasures + \numbColErasures \leq n-k$ in the sum-rank metric, where $n$ is the code length and $k$ is the code dimension.
The proposed decoder requires at most $\OCompl{n^2}$ operations in $\Fqm$ and, up to our knowledge, is the first scheme for \ac{LRS} codes capable of correcting both errors and erasures in the sum-rank metric.
We showed how the proposed decoder can be used for error control in noncoherent multishot network coding.
Future work will include error-erasure decoding of interleaved \ac{LRS} codes and consider the implications of errors and erasures in the \emph{skew metric}, which is isomorphic to the sum-rank metric.

\clearpage
\IEEEtriggeratref{17}
\bibliographystyle{IEEEtran}

\begin{thebibliography}{10}
\providecommand{\url}[1]{#1}
\csname url@samestyle\endcsname
\providecommand{\newblock}{\relax}
\providecommand{\bibinfo}[2]{#2}
\providecommand{\BIBentrySTDinterwordspacing}{\spaceskip=0pt\relax}
\providecommand{\BIBentryALTinterwordstretchfactor}{4}
\providecommand{\BIBentryALTinterwordspacing}{\spaceskip=\fontdimen2\font plus
\BIBentryALTinterwordstretchfactor\fontdimen3\font minus
  \fontdimen4\font\relax}
\providecommand{\BIBforeignlanguage}[2]{{%
\expandafter\ifx\csname l@#1\endcsname\relax
\typeout{** WARNING: IEEEtran.bst: No hyphenation pattern has been}%
\typeout{** loaded for the language `#1'. Using the pattern for}%
\typeout{** the default language instead.}%
\else
\language=\csname l@#1\endcsname
\fi
#2}}
\providecommand{\BIBdecl}{\relax}
\BIBdecl

\bibitem{lu2005unified}
H.-F. Lu and P.~V. Kumar, ``{A Unified Construction of Space-Time Codes with
  Optimal Rate-Diversity Tradeoff},'' \emph{IEEE Transactions on Information
  Theory}, vol.~51, no.~5, pp. 1709--1730, 2005.

\bibitem{nobrega2010multishot}
R.~W. N{\'o}brega and B.~F. Uch{\^o}a-Filho, ``{Multishot Codes for Network
  Coding using Rank-Metric Codes},'' in \emph{2010 Third IEEE International
  Workshop on Wireless Network Coding}.\hskip 1em plus 0.5em minus 0.4em\relax
  IEEE, 2010, pp. 1--6.

\bibitem{wachter2011partial}
A.~Wachter, V.~R. Sidorenko, M.~Bossert, and V.~V. Zyablov, ``{On (Partial)
  Unit Memory Codes Based on Gabidulin Codes},'' \emph{Problems of Information
  Transmission}, vol.~47, no.~2, pp. 117--129, 2011.

\bibitem{wachter2015convolutional}
A.~Wachter-Zeh, M.~Stinner, and V.~Sidorenko, ``{Convolutional Codes in Rank
  Metric with Application to Random Network Coding},'' \emph{IEEE Transactions
  on Information Theory}, vol.~61, no.~6, pp. 3199--3213, 2015.

\bibitem{napp2017mrd}
D.~Napp, R.~Pinto, J.~Rosenthal, and P.~Vettori, ``{MRD Rank Metric
  Convolutional Codes},'' in \emph{IEEE International Symposium on Information
  Theory (ISIT)}.\hskip 1em plus 0.5em minus 0.4em\relax IEEE, 2017, pp.
  2766--2770.

\bibitem{napp2018faster}
------, ``{Faster Decoding of Rank Metric Convolutional Codes},'' in \emph{23rd
  International Symposium on Mathematical Theory of Networks and Systems},
  2018, pp. 507--510.

\bibitem{byrne2021fundamental}
E.~Byrne, H.~Gluesing-Luerssen, and A.~Ravagnani, ``{Fundamental Properties of
  Sum-Rank-Metric Codes},'' \emph{IEEE Transactions on Information Theory},
  vol.~67, no.~10, pp. 6456--6475, 2021.

\bibitem{martinez2018skew}
U.~Mart{\'\i}nez-Pe{\~n}as, ``{Skew and Linearized Reed--Solomon Codes and
  Maximum Sum Rank Distance Codes over any Division Ring},'' \emph{Journal of
  Algebra}, vol. 504, pp. 587--612, 2018.

\bibitem{martinez2019reliable}
U.~Mart{\'\i}nez-Pe{\~n}as and F.~R. Kschischang, ``{Reliable and Secure
  Multishot Network Coding Using Linearized Reed--Solomon Codes},'' \emph{IEEE
  Transactions on Information Theory}, vol.~65, no.~8, pp. 4785--4803, 2019.

\bibitem{martinez2019Universal}
------, ``{Universal and Dynamic Locally Repairable Codes with Maximal
  Recoverability via Sum-Rank Codes},'' \emph{IEEE Transactions on Information
  Theory}, vol.~65, no.~12, pp. 7790--7805, 2019.

\bibitem{puchinger2020generic}
S.~Puchinger, J.~Renner, and J.~Rosenkilde, ``{Generic Decoding in the Sum-Rank
  Metric},'' in \emph{2020 IEEE International Symposium on Information Theory
  (ISIT)}.\hskip 1em plus 0.5em minus 0.4em\relax IEEE, 2020, pp. 54--59.

\bibitem{bartz2021decoding}
H.~Bartz and S.~Puchinger, ``{Decoding of Interleaved Linearized Reed--Solomon
  Codes with Applications to Network Coding},'' in \emph{IEEE International
  Symposium on Information Theory (ISIT)}.\hskip 1em plus 0.5em minus
  0.4em\relax IEEE, 2021, pp. 160--165.

\bibitem{bartz2022fast}
------, ``{Fast Decoding of Interleaved Linearized Reed–-Solomon Codes and
  Variants},'' \emph{submitted to: IEEE Transactions on Information Theory},
  2022, available at https://arxiv.org/abs/2201.01339.

\bibitem{hormann2021efficient}
F.~H{\"o}rmann and H.~Bartz, ``{Efficient Decoding of Folded Linearized
  Reed--Solomon Codes in the Sum-Rank Metric},'' in \emph{The Twelfth
  International Workshop on Coding and Cryptography (WCC)}, 2022, available at
  https://arxiv.org/abs/2109.14943.

\bibitem{GabidulinParamonovTretjakov_RankErasures_1991}
E.~M. Gabidulin, A.~V. Paramonov, and O.~V. Tretjakov, ``{Rank Errors and Rank
  Erasures Correction},'' in \emph{4th International Colloquium on Coding
  Theory}, 1991, pp. 11--19.

\bibitem{puchinger2021private}
S.~Puchinger and U.~Mart{\'\i}nez-Pe{\~n}as, Personal Communication, 2021, to
  appear in an extended version of arXiv:2109.09551.

\bibitem{richter2004ErrorAndErasure}
G.~Richter and S.~Plass, ``{Error and Erasure Decoding of Rank-Codes with a
  Modified Berlekamp--Massey Algorithm},'' in \emph{5th International ITG
  Conference on Source and Channel Coding (SCC), Erlangen}, 2004, pp. 203--210.

\bibitem{silva2009error}
D.~Silva, ``{Error Control for Network Coding},'' Ph.D. dissertation,
  University of Toronto, 2009.

\bibitem{silva_rank_metric_approach}
D.~Silva, F.~R. Kschischang, and R.~K{\"o}tter, ``{A Rank-Metric Approach to
  Error Control in Random Network Coding},'' \emph{IEEE Transactions on
  Information Theory}, vol.~54, no.~9, pp. 3951--3967, 2008.

\bibitem{lam1988vandermonde}
T.-Y. Lam and A.~Leroy, ``{Vandermonde and Wronskian Matrices over Division
  Rings},'' \emph{Journal of Algebra}, vol. 119, no.~2, pp. 308--336, 1988.

\bibitem{li2014transform}
W.~Li, V.~Sidorenko, and D.~Silva, ``{On Transform-Domain Error and Erasure
  Correction by Gabidulin Codes},'' \emph{Designs, Codes and Cryptography},
  vol.~73, no.~2, pp. 571--586, 2014.

\bibitem{leroy1995pseudolinear}
A.~Leroy, ``{Pseudo Linear Transformations and Evaluation in Ore Extensions},''
  \emph{Bulletin of the Belgian Mathematical Society-Simon Stevin}, vol.~2,
  no.~3, pp. 321--347, 1995.

\bibitem{martinez2019private}
U.~Mart{\'\i}nez-Pe{\~n}as, ``{Private Information Retrieval from Locally
  Repairable Databases with Colluding Servers},'' in \emph{IEEE International
  Symposium on Information Theory (ISIT)}.\hskip 1em plus 0.5em minus
  0.4em\relax IEEE, 2019, pp. 1057--1061.

\bibitem{caruso2019residues}
X.~Caruso, ``{Residues of Skew Rational Functions and Linearized Goppa
  Codes},'' \emph{arXiv preprint arXiv:1908.08430v1}, 2019.

\bibitem{caruso2021duals}
X.~Caruso and A.~Durand, ``{Duals of Linearized Reed--Solomon Codes},''
  \emph{arXiv preprint arXiv:2110.12675}, 2021.

\bibitem{gabidulin2008errorAndErasure}
E.~M. Gabidulin and N.~I. Pilipchuk, ``{Error and Erasure Correcting Algorithms
  for Rank Codes},'' \emph{Designs, Codes and Cryptography}, vol.~49, no. 1-3,
  pp. 105--122, 2008.

\bibitem{gabidulin1992AFastMatrixDecodingAlg}
E.~M. Gabidulin, ``{A Fast Matrix Decoding Algorithm for Rank-Error-Correcting
  Codes},'' in \emph{Algebraic Coding}.\hskip 1em plus 0.5em minus 0.4em\relax
  Springer-Verlag, 1992, pp. 126--133.

\bibitem{Sidorenko2011SkewFeedback}
V.~R. Sidorenko, L.~Jiang, and M.~Bossert, ``{Skew-Feedback Shift-Register
  Synthesis and Decoding Interleaved {G}abidulin Codes},'' \emph{IEEE
  Transactions on Information Theory}, vol.~57, no.~2, pp. 621--632, 2011.

\bibitem{Gabidulin_TheoryOfCodes_1985}
E.~M. Gabidulin, ``{Theory of Codes with Maximum Rank Distance},''
  \emph{Problems of Information Transmission}, vol.~21, no.~1, pp. 1--12, 1985.

\bibitem{berlekamp2015algebraic}
E.~R. Berlekamp, \emph{{Algebraic Coding Theory (revised edition)}}.\hskip 1em
  plus 0.5em minus 0.4em\relax World Scientific, 2015.

\bibitem{stein_sagemath}
{W.~A. Stein \emph{et~al.}}, {SageMath} {Software}, {http://www.sagemath.org}.

\bibitem{nobrega2009multishot}
R.~W. N{\'o}brega and B.~F. Uch{\^o}a-Filho, ``{Multishot Codes for Network
  Coding: Bounds and a Multilevel Construction},'' in \emph{2009 IEEE
  International Symposium on Information Theory (ISIT)}.\hskip 1em plus 0.5em
  minus 0.4em\relax IEEE, 2009, pp. 428--432.

\bibitem{bartz2017algebraic}
H.~Bartz, ``{Algebraic Decoding of Subspace and Rank-Metric Codes},'' Ph.D.
  dissertation, Technical University of Munich, 2017.

\bibitem{WeakKeysFL}
T.~Jerkovits and H.~Bartz, ``{Weak Keys in the Faure-Loidreau Cryptosystem},''
  in \emph{7th Code-Based Cryptography Workshop (CBC)}, Darmstadt, Germany,
  2019, pp. 102--114.

\bibitem{RandomizedGabDecoding}
J.~Renner, T.~Jerkovits, H.~Bartz, S.~Puchinger, P.~Loidreau, and
  A.~Wachter-Zeh, ``{Randomized Decoding of Gabidulin Codes Beyond the Unique
  Decoding Radius},'' in \emph{11th International Conference on Post-Quantum
  Cryptography (PQCrypto)}, Paris, France, 2020, pp. 3--19.

\end{thebibliography}

\end{document}